# Understanding deformation with high angular resolution electron backscatter diffraction (HR-EBSD)

T Ben Britton and J L R Hickey

Imperial College London, Department of Materials, Exhibition Road, London SW7 2AZ, Great Britain

E-mail: b.britton@imperial.ac.uk

**Abstract.** High angular resolution electron backscatter diffraction (HR-EBSD) affords an increase in angular resolution, as compared to 'conventional' Hough transform based EBSD, of two orders of magnitude, enabling measurements of relative misorientations of $1 \times 10^{-4}$ rads (~ 0.006°) and changes in (deviatoric) lattice strain with a precision of $1 \times 10^{-4}$. This is achieved through direct comparison of two or more diffraction patterns using sophisticated cross-correlation based image analysis routines. Image shifts between zone axes in the two-correlated diffraction pattern are measured with sub-pixel precision and this realises the ability to measure changes in interplanar angles and lattice orientation with a high degree of sensitivity. These shifts are linked to strains and lattice rotations through simple geometry. In this manuscript, we outline the basis of the technique and two case studies that highlight its potential to tackle real materials science challenges, such as deformation patterning in polycrystalline alloys.

## 1. Introduction

Electron diffraction offers an opportunity to study the internal structure of materials. In the scanning electron microscope (SEM), we can perform electron diffraction using backscattering electrons.

In electron backscatter diffraction (EBSD), these electrons enter the sample and scatter through electron-matter interactions [1]. If the sample is tilted to a high angle, a high fraction of these scattering electrons escape, and if their last scattering angle is favourably oriented with respect to the crystal lattice they will diffract according to Bragg's law [2]. In the kinematic approximation of diffraction, edges to Kikuchi bands are formed at an angle ± θ with respect to the perpendicular of the plane normal of the diffracting plane [3]. The EBSD pattern contains many bands, which are located on the phosphor screen as a function of the microscope settings (i.e., voltage), crystal orientation, lattice parameters, and the location of the imaging screen (i.e., the pattern centre) [4]. The geometry of the electron scattering process results in the formation of a direct projection of the diffracting lattice planes.

A change in band width (e.g., due to a change in a hydrostatic strain state) can occur due to the change in interplanar spacing affecting Bragg's law but the conventional EBSD geometry is relatively insensitive to this [5]. A change in interplanar angles will cause the bands to move across the diffracting pattern and this is more obvious and therefore measurable. For a deformation experiment, these changes can be induced by the presence of a (relative) deviatoric strain, and furthermore the bands may move across the screen due to the crystal rotating due to the presence and movement of dislocations during crystal slip. The presence of stored dislocations which result in local orientation gradients, can be linked through Nye's analysis [6] and therefore high precision measurements of spatial rotation gradients can be used to estimate the lower bound of the stored dislocation content [7, 8].

High resolution electron backscatter diffraction increases the relative angular resolution of the technique by two orders of magnitude, through comparison of two diffraction patterns to extract subtle changes. The relative merits of EBSD and HR-EBSD are summarised in Table 1.

**Table 1**. A relative comparison of conventional (i.e., Hough based) and high resolution (i.e., cross-correlation based) electron backscatter diffraction

|  | EBSD | HR-EBSD |
| --- | --- | --- |
| Absolute orientation | ~2° [9] | No |
| Misorientation | ~0.1 to 0.5° [10, 11] | ~0.006° ($1\times10^{-4}$ rads) [7] |
| GNDs @ 1 µm step | > $3\times10^{13}$ | > $3\times10^{11}$ |
| GNDs @ 100 nm step in lines / m² (b = 0.3 nm) [12, 13] | > $3\times10^{14}$ | > $3\times10^{12}$ |
| Relative elastic strain | No | Deviatoric strain $\pm 1\times10^{-4}$ |
| Relative residual stress (Type III – within grain) | No | Using anisotropic Hooke's law $\pm$ ~20 MPa (when E = 200 GPa) |
| Example tasks: | Microstructure, Texture, Grain size | Deformation i.e., elastic strain, misorientation and residual dislocation content |

Conventional, Hough based, EBSD uses band identification, through peak identification in Hough space, and subsequent indexing to identify the crystal orientation [4]. Work by Humphrey's indicates that the effective precision of this approach for misorientation analysis is ~ 0.5° (i.e., 0.0087 rads) [14]. The high resolution EBSD analysis, introduced principally by Wilkinson, Meaden and Dingley [7, 8] in 2006 directly compares the two diffraction patterns using cross correlation based image analysis to precisely measure the change in interplanar angles and results in an increase in measurement precision of two orders of magnitude, i.e., 0.0001 rads [7, 8].
These subtle shifts can be ascribed to lattice deformations resulting from a change in the lattice parameters due to a deviatoric change in lattice strain and therefore enable, using Hooke's law and a plane stress assumption, the residual stress variations between points within a single grain to be evaluated with a precision of ~20 MPa (for a material with a Young's modulus of 200 GPa).

The HR-EBSD technique has been widely applied across a range of semiconductor [7, 15], superconductor [16, 17], metals (e.g., [18-20]) and geological systems [21], often within the scope of understanding local changes in lattice parameter and especially within the field of deformation analysis.

In this manuscript, we will briefly introduce the key methodology of the technique. This will be followed by an example showing its efficacy at resolving high resolution orientation gradients in deformed copper and focus on studying the storage of geometrically necessary dislocations around an indent made in interstitial free (IF) steel.

## 2. Methodology
HR-EBSD involves analysis of diffraction patterns that are captured using the conventional EBSD set-up, with the camera inserted as for a texture or grain mapping experiment. The only differences are that the quality of the sample preparation must be higher (as surface scratches or deformation is more evident when the precision of the experiment is higher) and that the best quality patterns are captured. The quality of the patterns is important as signal to noise ratio and angle subtended per pixel (i.e., the pattern resolution) affects the resolution of the result [13, 22, 23]. HR-EBSD can be conducted both with so called 'fast' cameras or high binning modes of more 'sensitive' cameras [23], yet for the ultimate resolution often the images are not binned and often long (up to 1s) exposures with moderate probe currents (~10 nA) are required. The quality of the camera can affect the accruacy of the fitting of pattern shifts to strain, through errors in pattern centre and optical distortions [24]. However for relative (experimental pattern to experimental pattern) HR-EBSD approaches measurement of the pattern centre within 5 pixels [24] and typical optical distortions are reasonable [24].

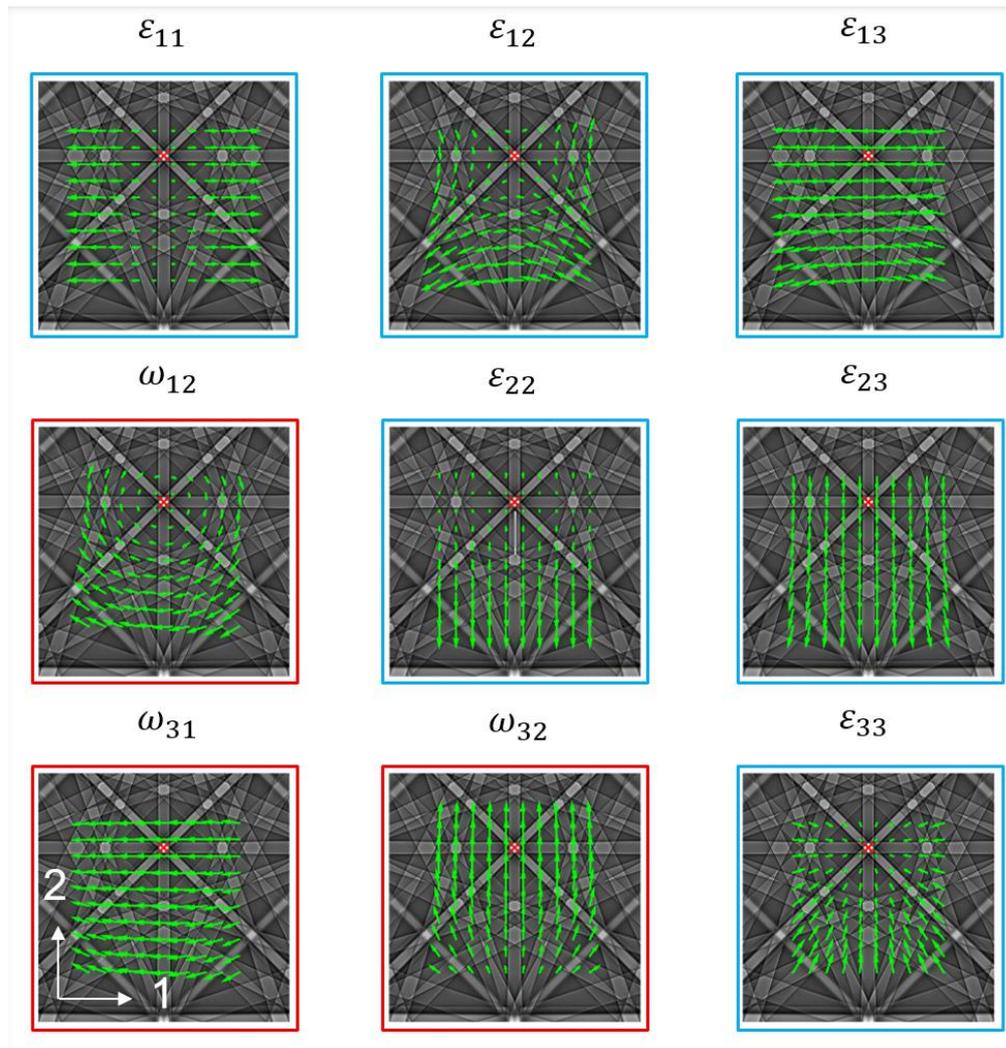

**Figure 1**. The effect of strain and rotation components on region of interest shifts, for deformations performed in the phosphor frame. Note the presence of the pattern centre, indicated as a red mark. The three normal strain components are not independent and so hydrostatic strains are not accessible with this approach.

Once an array of patterns is captured to disk, the processing is conducted offline using software algorithms. An EBSD map is segmented into individual grains containing regions of similar orientation and a reference point, ideally with no strain, but often just of high pattern quality, is selected. The reference point must be less than ~ 12° from the orientation of the test pattern, or there is insufficient overlap between test and reference for a reasonable cross-correlation result [25]. The HR-EBSD technique measures change in strain and lattice rotation with respect to the reference point, and so these may not be variations in absolute strain.

The reference pattern is sub-divided into many 'regions of interest' (ROI), which for a 1,000 x 1,000 pixel image are often 256 x 256 pixels in size [8] and this is repeated for a test pattern, extracted for a point within the same grain. At each ROI, the pattern is extracted and transferred to the Fourier domain for filtering and frequency based cross-correlation. The FFT based cross-correlation enables rapid calculation of the best translation vector between test and reference ROIs. If the translation of four or more well-spaced ROIs are measured well, then with knowledge of the pattern centre, eight components of the deformation tensor can be extracted. For the impact of strain and lattice rotation on the pattern, see Fig. 1. The ninth of deformation is the hydrostatic strain, which can be calculated only using an out of plane stress traction free boundary condition and Hooke's law.

Reasonable measurement of the translation shift is achieved using sub-pixel precision through up-sampling techniques (i.e., the centre of the cross-correlation peak is determined with sub-pixel accuracy as we assume that the shape of the peak is smooth – see [26] for more information).

Mapping strain and rotation within a map requires capturing of two patterns with the same geometry. Typically, EBSD maps are capturing using 'beam scanning' approaches where the electron beam is deflected and rastered across the sample. If the sample is flat and well aligned in the microscope (i.e. the sample plane normal is well known) then it is possible to construct a simple geometrical model where the beam shift can be corrected for (this is trivially accounted for in commercial software). This correction is important when the beam is shifted by a similar amplitude to the pattern shifts which arise due to strain or lattice rotation. Modern indirect detectors have a effective pixel size of ~20-40 μm, and therefore a uncertainties of in the prediction of the beam shift of 1 μm can result in systematic errors in the calculation of deformation tensors. Alignment of the sample can be assisted greatly through careful insertion into the SEM and presentation with respect to the beam, for instance using special sample holders (for example see page 85, chapter 2.3.3.2 of reference [27]). Note that for GND analysis (on which this paper focusses), beam shift errors have a less significant role as GNDs are calculated using local gradients between neighbour points).

Recently, it has been shown by both Maurice *et al.* [28] and Britton and Wilkinson [25] that the presence of a significant rotation gradient between test and reference pattern results in significant inaccuracy in the measurement of elastic strain, e.g., for metallic systems. Therefore, the algorithm has recently been updated to include a remapping step [25], whereby the pattern intensities of one pattern are remapped to better match the orientation of the other prior to extracting the strain components. The total contribution of the remapping and the shift measurements are combined to extract the deformation gradient tensor.

The deformation gradient tensor can be split into strain and rotation using either a finite or infinitesimal decomposition framework. In the case of small strains and small rotations these two are approximately equivalent, and for HR-EBSD, the finite framework is typically used for cases of large lattice rotation differences (as the strain gradients are typically small, due to the elastic limit in most materials).

The HR-EBSD pattern matching approach provides the best 'guess' to fit a model that maps the intensities of one pattern onto another, and this model is used to calculate the deformation gradient (see Fig. 2). This approach has been validated extensively using both experimental comparisons, e.g., through comparisons with NIST traceable Raman based stress measurements and AFM in Si [29, 30], comparison with modelling [15, 23, 32], comparison with X-ray based micro-Laue diffraction [32], and algorithm validation using dynamically simulated diffraction patterns [15, 22, 23].

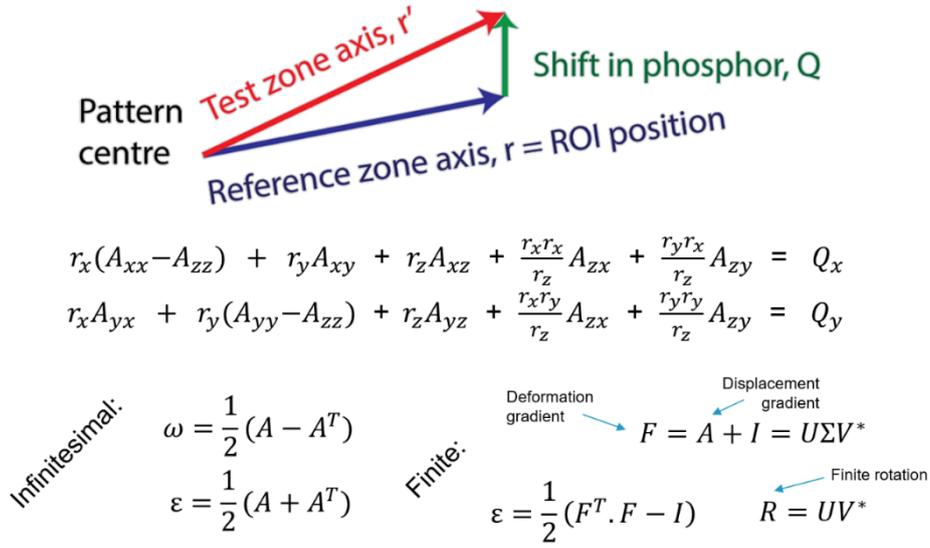

$$r_x(A_{xx}-A_{zz}) + r_y A_{xy} + r_z A_{xz} + \frac{r_x r_x}{r_z} A_{zx} + \frac{r_y r_x}{r_z} A_{zy} = Q_x$$

$$r_x A_{yx} + r_y(A_{yy}-A_{zz}) + r_z A_{yz} + \frac{r_x r_y}{r_z} A_{zx} + \frac{r_y r_y}{r_z} A_{zy} = Q_y$$

Infinitesimal:
$$\omega = \frac{1}{2}(A - A^T)$$
$$\varepsilon = \frac{1}{2}(A + A^T)$$

Finite:
$$\varepsilon = \frac{1}{2}(F^T \cdot F - I)$$

Deformation gradient: $F = A + I = U\Sigma V^*$ — Displacement gradient

Finite rotation: $R = UV^*$

**Figure 2.** Relationship between shifts in the phosphor and extraction of strain and lattice rotation for HR-EBSD.

One particular aspect of this work has been the generation of 'quality' metrics [26] that indicate when the HR-EBSD method fails. The mean angular error (MAE) describes the difference between the pattern shifts as measured and the predicted pattern shifts per the model, and a high value (i.e., greater than the measurement made) indicates the precision, and not accuracy, obtained. The (normalised) peak height describes the degree of matching for the cross-correlation peak, and it is often formatted as a geometric mean of all ROI peak heights normalised against auto correlation. Measurements using simulations have indicated that when this value falls below 0.3, the measurement is unlikely to be reasonable [32]. These two metrics can be used to rapidly screen data from large maps to remove suspect points, however further care must be taken when reporting stress variations (for example) as for example, not using a remapping step can result in artificially high stresses to be reported which are not physically reasonable [33].

Once the lattice rotations between test and reference are measured, it is possible to utilise Nye's framework [6] to estimate the geometrically necessary dislocation (GND) density and this has proved a very popular aspect of the technique [12, 13, 19, 34, 35]. In short, for each point within the orientation gradient map, a local gradient of the lattice rotation field is calculated and use to determine the lattice curvature (the adaptive kernel we use is described in [36]). As three rotation terms are measured in a 2D map, six components of lattice curvature are calculated. From Nye's formulation, the presence of GNDs can support these lattice curvatures, and it is a 'simple' task of solving the following equation for $\boldsymbol{\rho}$:

$$\boldsymbol{\xi}(6 \times 18) \cdot \boldsymbol{\rho}(18 \times 1) = \boldsymbol{\Lambda}(6 \times 1) \qquad (1)$$

where $\boldsymbol{\rho}$ are the curvature types, $\boldsymbol{\Lambda}$ are lattice curvatures, and $\boldsymbol{\xi}$ is the outer product of the Burgers vector and line direction for each dislocation type (the origin of this equation and its use for HR-EBSD measurements is explained in Wilkinson and Randman [12]).

For a metallic system, such as for FCC crystals such as Cu, the equation is underdetermined as there are many more potential dislocation types (12 {111} <110> edge, and 6 <110> screw [37]) than constraints from measurement of lattice curvature and therefore following the work of Wilkinson and Randman [12], we have chosen to find a potential solution that reduces the self-energy of the dislocations (accounting for screw and edge types, as well as changes in Burgers vector length) in a so called "L-1 minimisation" formulation. This solution is one of many solutions to the problem that necessarily supports the lattice curvature, but it may not be unique.

## 3. Example – Deformed Cu and geometrically necessary dislocations

For this example, we take a map from data reported elsewhere [18, 38] and use this to explore the relative merits of extracting GND content using Hough and HR-EBSD based approaches. There are two importance differences between Hough and HR-EBSD analysis: HR-EBSD has a higher angular misorientation precision and measures the misorientation about three independent axes (see Figure 2); the Hough based measurement of disorientation has a greater angular misorientation uncertainty, and as the misorientation angle between two points decreases the axis of misorientation is less well known [39, 40]. This makes the formation of a confident solution to equation 1 difficult (irrespective of the underdetermined nature of this equation for a metallic system).

The map, taken from a sample of uniaxially deformed Cu, was captured using 1x1 binning by Jun Jiang at Oxford University on a JEOL 6500F instrument with 1x1 binning and a Digiview II camera and OIM-DC v5 system. HR-EBSD analysis was performed using fifty 256 x 256 pixel ROI.

A comparison of the disorientation fields, using the same reference point between Hough and HR-EBSD techniques, is shown in Fig. 3. Qualitatively 2D field plots look identical between the techniques, yet a quantitative line scan indicates that the Hough based approach has a lower signal to noise ratio and the magnitude of these noise fluctuations varies from grain to grain.

These disorientation fields (Fig. 3) were input into the same GND code and the total GND density was calculated and they are reported in Fig. 4.

The presence of high frequency noise in the Hough based disorientation analysis results in an apparent increase in the GND density recovered, this is evident in a simple area fraction analysis (Fig. 4d). Furthermore, the HR-EBSD analysis is likely to be relatively isotropic in terms of noise floor with regards to crystal orientation, whereas the Hough base approach seems to show and indicate that there are some grains and some grain boundaries where the apparent GND density is higher. This may be problematic if conclusions regarding the structure and accumulation of GNDs in particular grains or near particular grain boundaries are of interest to the investigator.

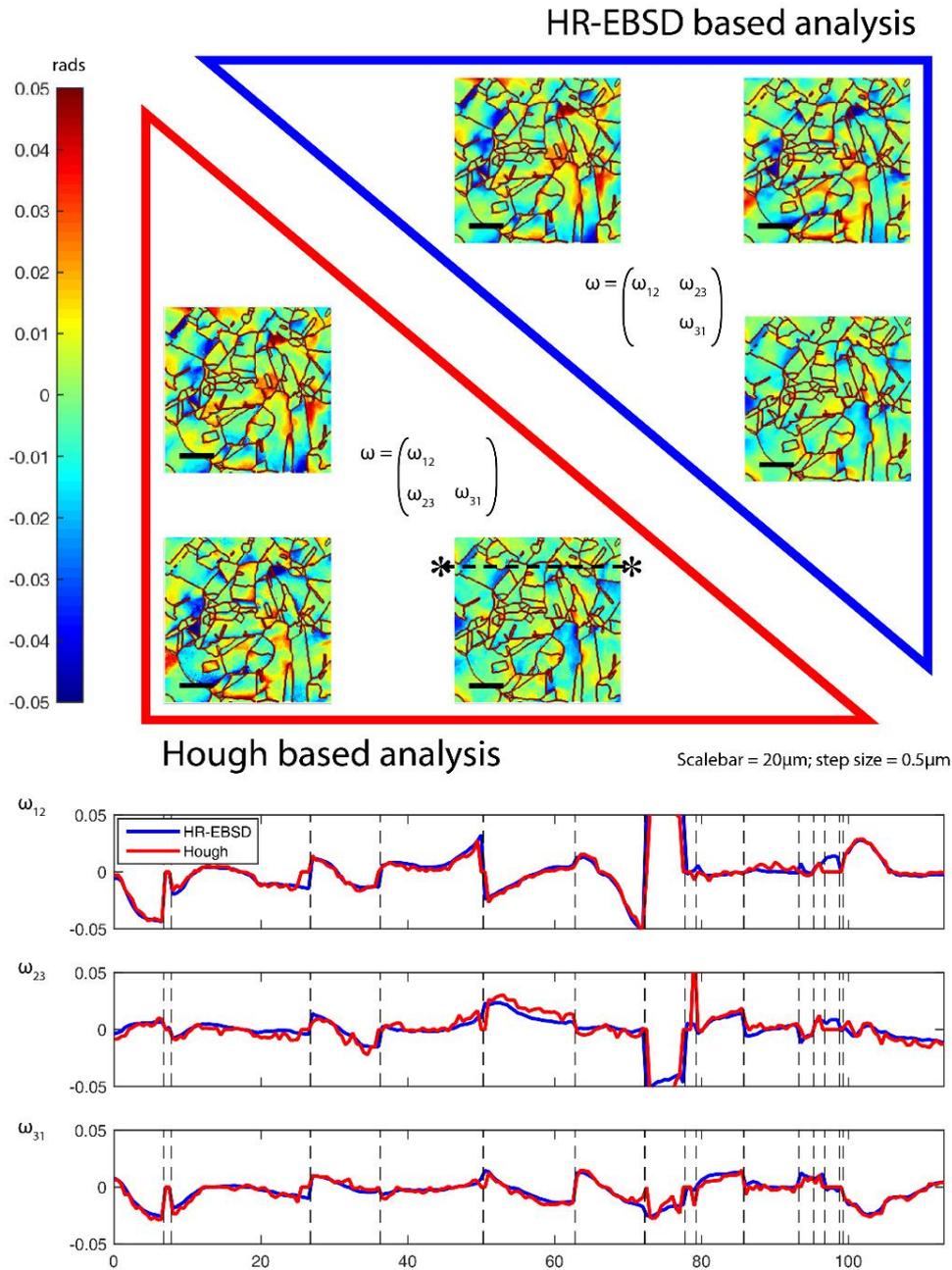

**Figure 3.** Lattice disorientation maps measured with HR-EBSD and Hough based methods for copper deformed to 6 % in strain. Vertical dashed lines represent grain boundaries. There is good agreement between both measurements and lattice disorientation varies for points within each grain, and the disorientation is often found to change significantly near grain boundaries. For these maps, the X1 axis points left to right, the X2 axis points down the field. (We thank Dr J. Jiang and Mr C. Zhou for their assistance in generating this map.)

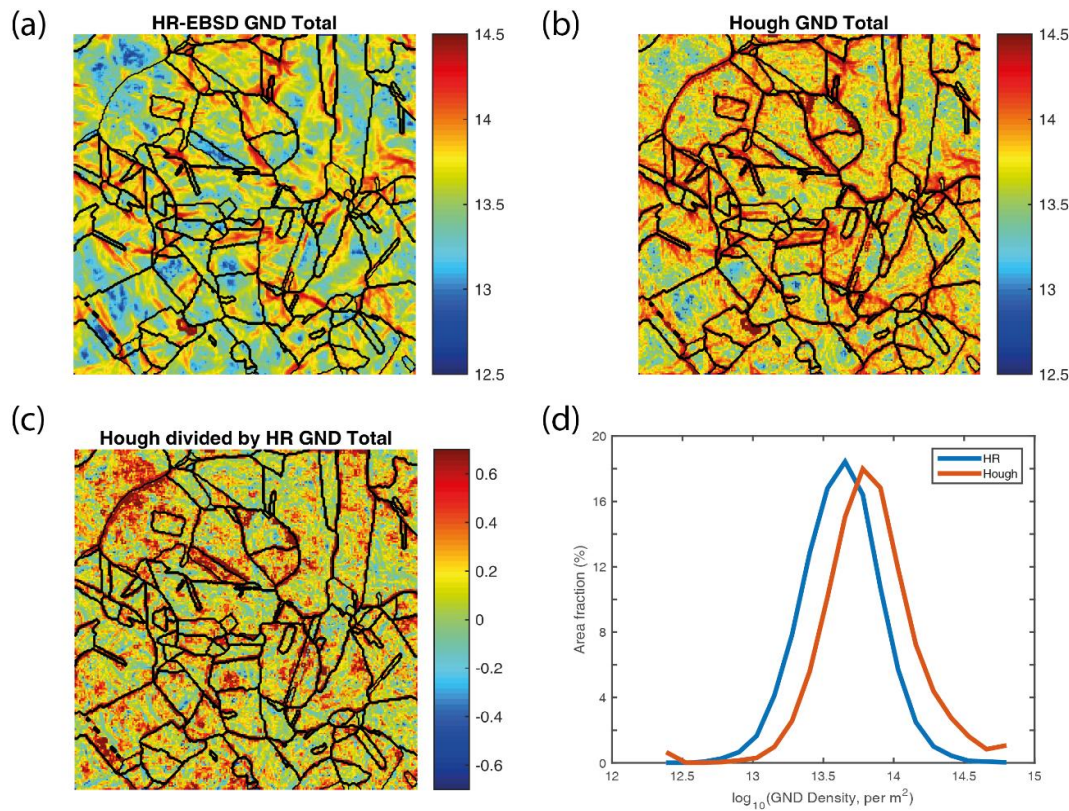

**Figure 4**. Comparison of GND measurements with Hough and HR-EBSD measurement of lattice curvature. (a and b) GND density distribution measured as $\log_{10}$(dislocations per $m^2$) for HR-EBSD and Hough based EBSD calculations respectively; (c) the Hough derived field divided pixel wise by the HR-EBSD field, presented on a $\log_{10}$ scale; (d) a histogram of the resultant plots. (a-c) represent areas that are 112.5 x 112.5 μm$^2$. (We thank Dr J. Jiang and Mr C. Zhou for their assistance in generating this map.)

## 4. Example – Indents in interstitial free steel

For this example, we have indented an interstitial free steel sample using a macro-indent and focussed on exploring the GND content surrounding the indenter impression. The indent spanned multiple grains and this study is interesting with respect to an understanding of how deformation patterns around micro-hardness measurements, which are widely used in industry as a measure of microstructure condition and quality assessment.

The sample was prepared as a 10 mm x 10 mm x 0.8 mm token from a rolled interstitial free steel sheet. The sample was heat-treated for 2 hours at 700 °C and quenched in room temperature water to recrystallise the microstructure. The RD-TD surface of the sample was subsequently polished to a 4,000 SiC finish. Final preparation of the sample surface was completed using a Gatan PECSII argon ion polisher using a three-step process, adjusting in voltage and/or angle with each 2 hr step: 8 kV at 5°, 3 kV at 5°, 3 kV at 1°.

A Vickers hardness indent was made into the surface of the sample within the region that was ion polished. The load applied to the indenter was 10 g for 5 s.

For the EBSD, a FEI Quanta 650g FEG-SEM coupled with a Bruker eFlash HR EBSD detector was used. The working and detector distances were set at 16.4 mm and 15.7 mm respectively. The sample and detector were tilted 70° and 10° respectively. EBSD patterns were captured at full resolution (1,200 x 1,600 pixels) at an exposure time of 0.6 s, probe current of ~ 10 nA, and the step size was 0.3 µm.

HR-EBSD analysis was performed with fifty 256 x 256 pixel ROI, with pattern remapping. GND evaluation was performed using Nye's analysis. As IF steel is ferritic, the <111> {110} slip systems were used (as per Wilkinson and Randman [12]).

The mean angular error (MAE) and cross-correlation peak heights (PHs) maps for the second cross-correlation pass are shown in Fig. 5. As the indent is broad, patterns were obtained from inside the indentation crater, yet the focus of this work is to explore the GND density surrounding the indenter impression and so these were filtered out of the GND analysis. These quality maps indicate that there is reasonable pattern quality and HR-EBSD analysis for the majority of grains, and a filter of PH > 0.3 and MAE < 5 x$10^{-3}$ was used. We note that there was a reduction in PH specifically in regions that were surrounding the indent and near grain boundaries, most likely to the presence of significant strain gradients and/or pattern overlap which affects diffraction within the interaction volume result in blurring of the diffraction pattern.

The HR-EBSD rotation gradient fields, combined from passes 1 and 2 of the cross-correlation process, were used to generate a map of the total GND density and this is shown in Fig. 6.

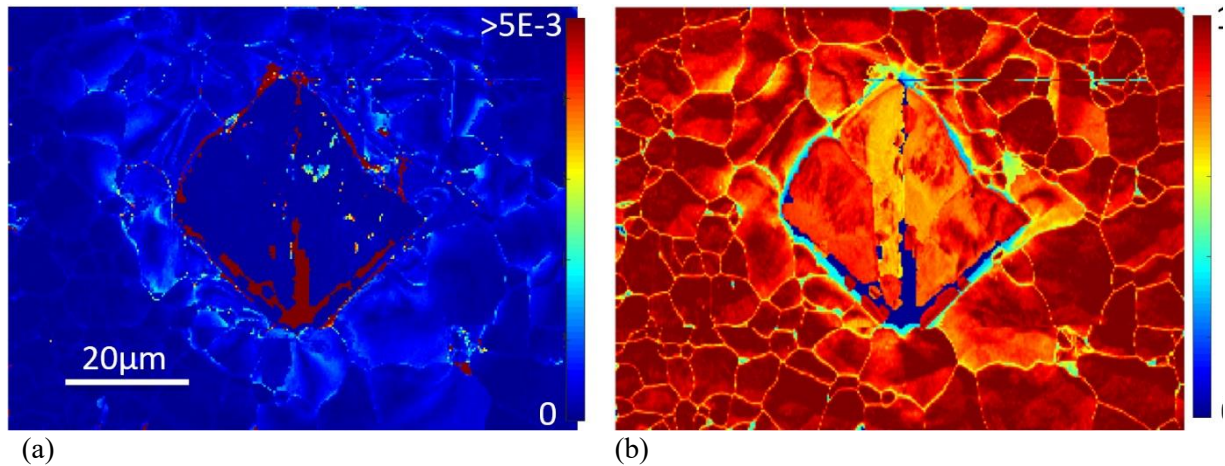

(a) (b)

**Figure 5.** Quality maps for the 2nd pass of the cross-correlation showing: a) mean angular error (in radians); b) Normalised cross-correlation peak heights.

In this map, we see that there are several notable features that highlight the heterogeneous nature of plastic slip in metallic systems, even when deformation is fiercely localised around an indenter.

A first order evaluation highlights that the GND density extends away from the indenter impression. Evidence of the plastic zone field (of which the GND density is only a part) indicates that plasticity is different within each grain, which is not surprising given the nature of crystal plasticity in a polycrystalline aggregate.

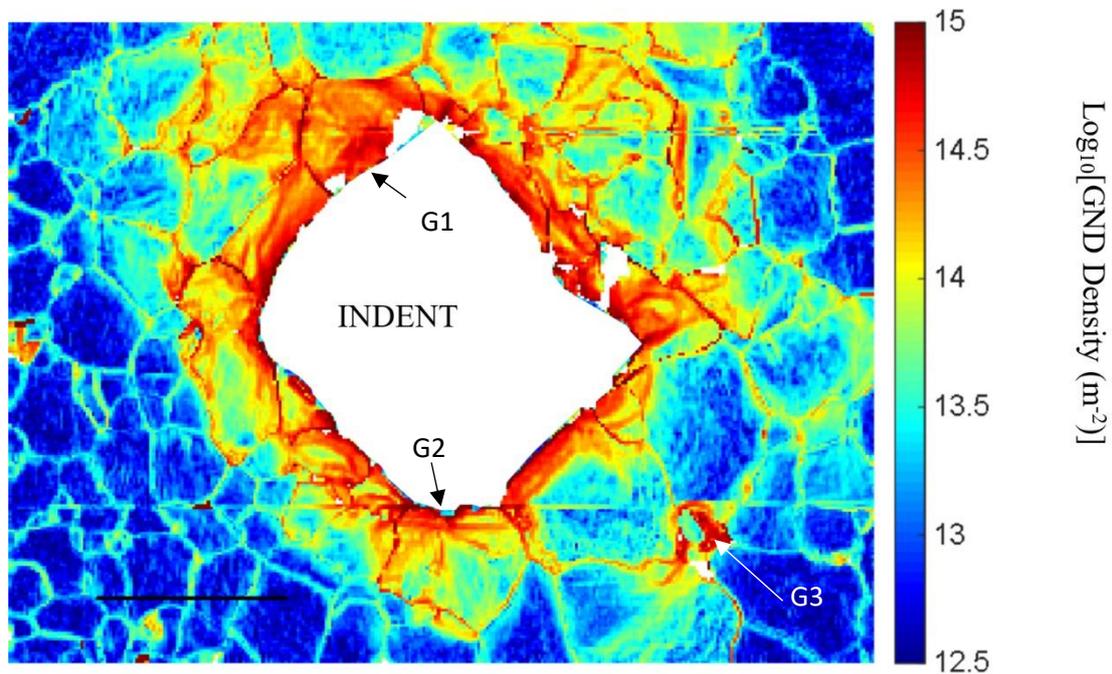

**Figure 6.** GND field around the indent. The length of the scale-bar is 20 µm.

The striking details of this heterogeneous plastic deformation are brought out in this micrograph, for instance there is one grain towards the top of the indent (G1), which has stored a significant amount of GND density in a smooth and continuous field; where as a grain towards the bottom of the indent (G2) has 'plumes' of high GND density extending from the vertex of the indenter and there are many other grains where only half of the grain near the indenter have significant GND density. Furthermore, while most of the high GND density is localised towards the impression edge, there are some regions 'one grain further in' (bottom right of the field, G3) which have significant accumulation of GND density towards the grain boundary region and there is a region clear of GND density nearer the impression.

Finally, compatibility and equilibrium conditions require that grains must 'co-deform' to accommodate strain (or cracks will form) and there is significant evidence of this in the GND field. The vein structure of the GND field is often modulated by the presence of grain boundaries, for instance where the particular stress state imposed during indentation enables one grain to freely plastically deform and store lattice strain gradients, and its neighbour does not. However, there are occasions where the grain boundaries almost seem 'invisible' to the GND structure, as veins of high GND content can extend across the grain boundary region, most likely due to local compatibility requirements.

We also note that there are two horizontal 'flashes' across the map at the bottom and top of the indenter impression which are due to stochastic image drift in the SEM, these are aesthetically displeasing but do not detract from our ability to interpret this data rich micrograph.

Due to the out of plane displacement, associated with pile-up or sink-in around the indent for this example, the exact magnitude of the GND field must be taken with care, as significant changes in surface roughness may impact the spatial field used to evaluate the curvature terms (i.e. we assume that the out of plane displacement is zero in formulating the curvature tensor). The impact of this uncertainty could be addressed using atomic force microscopy (AFM).

## 5. Discussion
HR-EBSD affords an increase in angular resolution of two orders of magnitude as compared to conventional Hough based EBSD. This increase in resolution enables probing of lattice strain variations, which can be used to calculate intragranular residual stress variations, and local orientation changes.

This is well suited for studies of deformed materials, such as the metallic examples shown here. This manuscript has focussed on GND evaluation, as this is a popular area of study and there have been examples of recent work that focus on the elastic strain measurement aspect of the technique, such as a review in JOM [26].

The HR-EBSD technique is sufficiently mature and robust that there are now multiple groups using the technique in industrial and academic contexts.

In this manuscript, we have outlined two case studies in deformed copper and iron that indicate the potential of the technique to probe the intricate nature of stored dislocation fields in deformed metals.

In the copper example, we highlight that an increased angular resolution, with explicit measurement of the misorientation axis, affords a greater reliability in the calculation of residual deformation content. Often this is more crudely reported through use of the kernel average misorientation (KAM), a step size independent metric of the local orientation gradient, and yet as EBSD provides knowledge of the crystal orientation, it is often possible to provide an educated guess at the potentially contributing slip systems, GND density maps can be easily calculated (and this has been extensively demonstrated by groups worldwide, including excellent work for example by Pantleon and co-workers [39, 40], as well as colleagues at the MPIE in Düsseldorf [43-45]). The absolute different between the two approaches when applied to the same sample is relatively small (a factor of 1.4x), and yet the standard deviation and localised variation is perhaps more significant.

The IF steel indenter impression highlights the potential of this technique to improve and further understand existing metallographic characterisation techniques. Observations of a component of the surface plastic zone indicates that crystal slip remains significantly heterogeneous, even where there is significantly localised plasticity. The presence of different GND density structures within different grains highlights that understanding localised plasticity around indents remains challenging, as there is an interplay of the local crystal orientation, the local grain neighbourhood, and the imposed strain gradient due to the indentation process.

Perhaps of wider interest, broadly we observe similar trends in the dislocation structures in these metallic systems across multiple grains where deformation in polycrystalline aggregates results in fierce localisation of stored dislocation content in some, but not all, grains and near some, but not all, grain boundaries and triple junctions. Understanding how these structures form and where localisation occurs in constrained polycrystalline samples is critical to develop a deeper understanding of deformation and failure of materials, and we hope that HR-EBSD is a valuable tool in the materials characterisation toolbox that will assist us on this journey.

## 6. Conclusion

We have reviewed the HR-EBSD methodology and outline key steps in direct pattern analysis to capture subtle changes in interplanar angle and lattice disorientation. The two examples presented here highlight the potential for HR-EBSD to add value to the materials characterisation community, affording an extra tool to characterise local structure with high precision within the scanning electron microscopy. The Cu example highlights how the increased angular resolution of HR-EBSD improves our confidence in probing local orientation gradients that can be interpreted in terms of stored GND content. We find that there are systematic variations in the GND density fields recovered, as well as localised regions where the HR-EBSD and Hough based approaches diverge significantly (such as near certain boundary types). Assessment of the GND content around an indent in IF steel highlights the potential for HR-EBSD to assist our understanding of deformation during metallographic tests and to relate this to the dominant crystal plasticity mechanisms at work.

**Acknowledgements**
We would like to thank Prof Angus Wilkinson for long standing and productive collaborations on the HR-EBSD technique. We acknowledge that Cu data was obtained by Dr J. Jiang during his PhD work at Oxford. We thank both C. Zhou and Dr J. Jiang for assistance with the initial code used to analyse the Cu GND data. TBB acknowledges funding from the Royal Academy of Engineering and funding from the AVIC-BIAM centre at Imperial College. TBB and JH both acknowledge funding from the Shell UTC in Advanced Interfaces in Materials Science. We would like to thank the reviewers for their useful comments which have improved the quality of this manuscript.